\input psfig.sty
\documentstyle{cupconf}

\ifoldfss
\else
  \ifnfssone
    \newmathalphabet{\mathit}
      \addtoversion{normal}{\mathit}{cmr}{m}{it}
      \addtoversion{bold}{\mathit}{cmr}{bx}{it}
    \newmathalphabet{\mathcal}
      \addtoversion{normal}{\mathcal}{cmsy}{m}{n}
    \else
    \ifnfsstwo
    \fi
  \fi
\fi

\def\la{\mathrel{\hbox{\rlap{\hbox{\lower4pt\hbox{$\sim$}}}\hbox{$<$}}}}
\def\ga{\mathrel{\hbox{\rlap{\hbox{\lower4pt\hbox{$\sim$}}}\hbox{$>$}}}}
\def\etal{\mbox{\it et al.}}
\font\cmss=cmss10 scaled 1200

\def\kms    {\ifmmode{{\rm ~km~s}^{-1}}\else{~km~s$^{-1}$}\fi}

\def\hexnumber#1{\ifcase#1 0\or1\or2\or3\or4\or5\or6\or7\or8\or9\or
 A\or B\or C\or D\or E\or F\fi }

\makeatletter
\ifx\CUP@mtlplain@loaded\undefined
\else
\fi
\makeatother
 \makeatletter
 \ifx\CUP@mtlplain@loaded\undefined
   \font\tenbmi=cmmib10 at 10pt
   \font\sevenbmi=cmmib10 at 7pt
   \font\fivebmi=cmmib10 at 5pt

   \newfam\bmifam
   \textfont\bmifam=\tenbmi
   \scriptfont\bmifam=\sevenbmi
   \scriptscriptfont\bmifam=\fivebmi
   
 \fi
 \makeatother
\ifnfsstwo

\fi
\ifnfssone

\fi
\ifoldfss

\fi

\mathchardef\varLambda="0103

\makeatletter
\ifx\CUP@mtlplain@loaded\undefined
\else
\fi
\makeatother
\makeatletter
\ifx\CUP@mtlplain@loaded\undefined
  \font\tenbms=cmbsy10
  \font\sevenbms=cmbsy10 at 7pt
  \font\fivebms=cmbsy10 at 5pt
  \newfam\bmsfam
  \textfont\bmsfam=\tenbms
  \scriptfont\bmsfam=\sevenbms
  \scriptscriptfont\bmsfam=\fivebms

  \edef\bsy@{\hexnumber\bmsfam}
  \mathchardef\bnabla="0\bsy@72
\fi
\makeatother
\def\etal{\mbox{\it et al.}}

\title[ ]{Accretion and Outflow in the Circinus AGN}

\author[ ]{Lincoln J. Greenhill$^{(1)}$}

\affiliation{$^{(1)}$ Harvard-Smithsonian Center for Astrophysics}

\begin{document}
\ifnfssone
\else
  \ifnfsstwo
  \else
    \ifoldfss
      \let\mathcal\cal
      \let\mathrm\rm
      \let\mathsf\sf
    \fi
  \fi
\fi

\maketitle

\begin{abstract}

The first VLBI images of H$_2$O maser emission in the Circinus Galaxy AGN show
both an accretion disk and outflow 0.1 to 1 pc from the central engine.  The
maser traces a warped, edge-on accretion disk between radii of about 0.1 and
0.4 pc that is bound by a $1.3\times10^6$ M$_\odot$ central mass.  The rotation
curve is somewhat flatter than a Keplerian rotation law and is consistent with
a disk mass on the order of $10^5$ M$_\odot$. Clumpy substructure may display a
kinematic signature consistent with spiral arms. Away from the disk, a second
population of water masers traces broadly a bipolar, wide-angle outflow that
contains (bullet-like) clumps ejected from a region centered on the central
engine and $<0.1$ pc in radius.  Out to a radius of $\sim 0.4$ pc, the warp of
the accretion disk appears to channel the outflow, the orientation of which on
the sky coincides with the orientation of the known kiloparsec-scale ionization
cone.  Beyond this radius, the flow crosses the disk and truncates it.
This suggests that the current accretion event and associated 
disk-outflow geometry
have a lifetime on the order of $10^7$ years.  

\end{abstract}

\firstsection 
\section{Introduction}

The Circinus galaxy is one of the nearest Seyfert\,II galaxies that hosts 
an extragalactic H$_2$O maser. (Freeman \etal\ 1977 estimate a distance of
4~Mpc.) The $\sim 10^{42}$ erg s$^{-1}$ (2-10 keV) central engine is obscured
at energies below 10 keV by a large gas column, n$_{\rm H}\sim4\times10^{24}$
cm$^{-2}$ (\cite{matt99}).  The X-ray spectrum also exhibits a prominent Fe
K$\alpha$ line that probably comprises  reflected radiation (\cite{matt96}). 
These characteristics suggest some  similarity to another maser-host galaxy,
NGC\,1068.

The Circinus AGN exhibits an ionization cone (\cite{marconi94}) within which
Veilleux
\& Bland-Hawthorn (1997) observed linear optical filaments and compact knots
reminiscent of Herbig-Haro objects. The cone opening angle is at least
$90^\circ$, at a mean position angle (PA) of roughly $290^\circ$. The AGN also
drives a kpc-scale nuclear outflow betrayed by bipolar radio lobes
(\cite{elmouttie98}) at PA $\sim 295^\circ$, which are largely aligned
with the minor axis the galactic H\,I disk (\cite{jones99}) and a nuclear
$^{12}$CO ring (\cite{curran98}).

There is strong evidence that at least several extragalactic H$_2$O masers
trace edge-on accretion disks bound by central engines $\ga 10^6$ M$_\odot$.
These masers display (1) roughly linear structure in projection on the sky, (2)
emission symmetrically bracketting the galactic systemic velocities, and (3)
declining rotation curves that indicate more or less Keplerian differential
rotation.  Discovery of similar  symmetry in the Circinus H$_2$O maser spectrum
(Nakai\ \etal\  1995; Greenhill\ \etal\  1997) led to speculation that there too the
masers trace an accretion disk (\cite{greenhill97}). Though the observation of
an accretion disk in Circinus confirms that prediction, the discovery of
qualitatively new, nondisk emission is equally interesting.  In NGC\,1068,
disk emission is accompanied by secondary maser emission $\gg1$ pc downstream
in a jet, where material in the narrow line region deflects it
(\cite{gallimore96}). However, the nondisk emission in Circinus 
lies $<1$ pc from the central engine and is more closely related to it. 

\section{Observations and Data}

We observed the Circinus H$_2$O maser three times for 18 hours each in 1997
(June \& July) and 1998 (June) with four stations of the Australia Telescope
Long Baseline Array, obtaining a half-power beamwidth of about $2\times4$
milliarcseconds (mas). We recorded two 16 MHz bandpasses ($\sim 215\kms$) at
each station and correlated the data with the S2 processor in Marsfield, NSW .
Following Fourier inversion of the correlation functions, each spectral channel
corresponded to $\sim 0.21$\kms, which provided at least three channels
across the half-power full-width of each spectral component.

We calibrated amplitudes, delays, and phases with standard VLBI techniques and
the AIPS package.  When possible we fitted a time-series of {\it total-power}
spectra for each station to a calibrated template spectrum and computed
multiplicative gain factors that would calibrate visibility amplitudes in units
of Janskys.  This calibration corrected for variations in antenna gain and in
atmospheric attenuation and emission, with a $<10\%$ relative accuracy
(station-to-station, moment-to-moment) and 30\% accuracy in absolute terms.
When during one 1997 track the total-power signal was too weak for template
fitting, we adjusted the gains to maintain constant peak visibility amplitude
in the {\it cross-power} spectra. We used scans of 0537-448, 1144-379, 1424-418,
and 1921-293 to estimate (and correct for) 
instrumental delays and fringe rates at
better than the 2 ns and 2 mHz level (1998) or 5 ns and 5 mHz level (1997). We
also estimated the stations' complex bandpass responses from observation
1424-418 and 1921-293.

We estimated a new astrometric position for the masers by analyzing the time
variation in fringe rate for the 565\kms~line. (We use the radio astronomical
definition of Doppler shift.)  The new position,
$\alpha_{2000}=14^h13^m09\rlap{.}^s95\pm0\rlap{}^s.02$,
$\delta_{2000}=-65^\circ20'21\rlap{.}''2\pm0\rlap{.}''1$, is the most accurate
estimate so far for the AGN.

We self-calibrated the emission within a few \kms~of 565\kms, applied the
calibration to each spectral channel, and created deconvolved synthesis images.
The noise in the images ($1\sigma$) was 0.025 - 0.045 Jy, depending on the epoch
and channel; the 1998 June observations were the most sensitive and best
calibrated, because of equipment upgrades and a (coincidental) 40 Jy flare that
made possible particularly accurate amplitude and phase calibrations. 
We fitted 2-D Gaussian model
brightness distributions to each emission component in the deconvolved images
that was stronger than $6\sigma$. 

At each epoch, the emission near 565\kms~was strongest.  However, because
spectral lines come and go from epoch to epoch, we superposed maps of the
emission component positions from each epoch, so as to trace the underlying
dense molecular gas as completely as possible with the available data
(Figure\,1). We registered the maps with 0.2 mas accuracy by comparing the
measured positions of the lines at 472 and 526\kms. (To achieve the stated
noise levels, the 1997 data were smoothed using a moving average of up to six
frequency channels, which ``blurs'' the map in crowded fields. The 1998 data
were Hanning smoothed.)

\section{Results}

The maser emission comprises complexes of spectral  lines that are redshifted
and blueshifted with respect to the systemic velocity of the galaxy.  In 1998
June, we detected emission between 214.8\kms~and 676.5\kms, a broader interval
than the roughly 250 to 650\kms~range reported by Nakai\ \etal\  (1995) and
Greenhill\ \etal\  (1997) in {\it total-power} observations.  The mean
velocity, $446\kms$, is consistent with the systemic H\,I velocity of the
galaxy, $438\pm2$\kms~(\cite{freeman77}), though emission on scales $>100$ pc
dominates measurement of the latter.

At each epoch, the sky distribution of maser emission may be divided into three
populations (see Figure\,1): (1) a thin, gently curved, {\cmss S}-shaped locus
of highly redshifted and blueshifted emission arcs to the southwest and
northeast, respectively (aka ``high-velocity'' emission),  (2) emission close
to the nominal systemic velocity of the galaxy (aka ``low-velocity'' emission)
that lies between the high-velocity arcs,  and (3) emission that is modestly
Doppler shifted and broadly distributed in knots that lie north and
west (redshifted) and south and east (blueshifted) of the low-velocity emission.

\subsection{The Warped Disk}

To the position and velocity data for high and low-velocity masers, we have
fitted a model edge-on disk with smoothly varying position angle as a function
of radius (Figures\,2 \& 3). The disk outer radius is $\sim 0\rlap{.}''02$ (0.4 pc)
and the thickness is $< 0\rlap{.}''002$ thick.   The orbital speed along the
inner edge of the disk, at a radius of $\sim 0.1$ pc, is 237\kms.  The mass
enclosed is $1.3\pm0.1\times10^6$ M$_\odot$, assuming circular motion, and the
model systemic velocity is $451\pm10$\kms.  (Uncertainties reflect formal 
errors.)  Among the redshifted high-velocity masers,
the peak rotation velocity as a function of impact parameter from the dynamical
center, $b$, traces a rotation curve that declines as approximately
$b^{-0.4\pm0.05}$ (Figure\,3), from which we infer a disk mass on the order of
$10^5$ M$_\odot$ between 0.1 and 0.4 pc.  The disk is edge-on at a radius of
0.1 pc, but if the disk becomes inclined toward the outer radius, then the disk
mass would be larger.

The model derives its strongest support from the following observations: (1)
the sky distribution of high-velocity masers is highly elongated, roughly
symmetric, and perpendicular to the approximate axis of the known ionization
cone, (2) the apparent rotation curve is nearly Keplerian, (3) the most highly
blue-shifted emission, innermost redshifted emission, and the low-velocity
emission lie along a line on the sky, and (4) a single position-velocity
gradient extends from the high-velocity blueshifted emission through the
low-velocity emission to the high-velocity redshifted emission.   The latter
two observations are strong signatures of an edge-on disk in which the
low-velocity emission lies close to the inner radius of the disk, as it does in
NGC\,4258 (\cite{miyoshi95}).

Together with the shallow rotation curve, apparent clumpy substructure 
suggests that self-gravity is important in the disk. Estimates of the Toomre
$Q$-parameter suggest that the disk is at best only marginally stable, due in
part to its slow rotation speed (cf Maoz (1995) in discussion of NGC\,4258).  
At the inner radius, where $\Omega$ is greatest, $Q\sim0.4\rho_9^{-1}
h_{-2}^{-1}$, where $\rho_9$ is H$_2$ particle density in units of $10^9$
cm$^{-3}$ and $h_{-2}$ is the disk thickness in units of 0.01 pc.  ($Q=c_s\Omega /
\pi G \Sigma$, where $c_s$ is the speed of sound, $\sim 1$\kms~at 400 K,
$\Omega$ is the angular rotation speed, and $\Sigma$ is the surface density.) 
Furthermore, the kinematic signature of at least two clumps among the redshifted
high-velocity masers may indicate that spiral arms are present, specifically, 
the decline in line-of-sight velocity with increasing impact parameter.  We
infer that these clumps are elongated along the line of sight and cross the disk
midline (i.e., the diameter perpendicular to the line of sight) at an angle. 
The angle is roughly the same in each case, which is suggestive of 
ordered structure, though the putative spiral may be fragmented.

\subsection{The Outflow}

A flow from the vicinity of the central engine is traced by: (1)
emission between 300 and 450\kms~south and east of the dynamical center of the
disk, and (2) emission between 450 and 580\kms, north and west of the dynamical
center.  This model demands that individual clumps are sufficiently luminous to
be observable despite their great distance, but this is reasonable. A
typical maser in the putative outflow is 0.1--0.5 Jy at a distance of 4 Mpc
and would appear to be $2-8\times10^4$ Jy were it at the distance of the W49N
star forming region, which is not unusual (\cite{liljestrom89}).  In contrast,
it is unlikely that the maser emission arises in the accretion disk because of
the irregular distribution of line-of-sight velocities on the sky, and because
were the emission to arise in a disk, it would not originate where amplification
paths (as dictated by geometry) are optimal. 

The segregation of red and blueshifted outflowing clumps on the sky suggests
that the flow is inclined with respect to the line of sight, while the
distribution on the sky suggests a wide-angle outflow.  However, our census of
outflowing clumps is incomplete because the inter-clump medium is probably
ionized, as in narrow-line regions, and  readily able to obscure emission from
the far side of the outflow.  (The known emission is close to our detection
limit and an optical depth of a few requires only readily attainable 
emission measures of
$5\times 10^9$ pc cm$^{-6}$.) Because H$_2$O maser emission requires H$_2$
densities $\sim 10^9$ cm$^{-3}$, the observed clumps are probably high-density
``bullets'' immersed in a thinner medium.  Comparable interclump densities
would imply a prohibitively large mass loss rate ($\gg 10$
M$_\odot$\,yr$^{-1}$) and unusually small accretion efficiency.

In principle the outflowing clumps may be ejected ballistically from the
vicinity of the central engine (at radii $<0.1$ pc) or shorn from the observed
molecular disk (e.g., \cite{emmering92};
Kartje, K\"onigl, \& Elitzur 1999). In the absence of acceleration or proper
motion data, ejection may be the preferred (and simplest) model for three
reasons.  First, some clumps lie close in projection to the rotation axis of the
accretion disk across  a range of impact parameters (i.e., from the disk inner
radius to at least twice the disk outer radius) and velocities.  Second, the
line-of-sight velocity and position data do not betray rotation.  Third, the
physical conditions in the maser clumps are consistent with adiabatic expansion
from conditions estimated for broad line clouds (i.e., densities of $10^{10}$
- 10$^{12}$ cm$^{-3}$ and temperatures of $10^4$ K, as in \cite{brotherton94}). 

The disk structure and rotation curve dissolve at a radius of $\sim 0.4$ pc,
where the wind appears to disrupt the accretion disk, whose density presumably
declines with increasing radius.  In the model of Neufeld, Maloney, \& Conger
(1994) maser excitation depends on irradiation of molecular gas by hard X-rays.
The proposition that the disk extends no farther than 0.4 pc from the central
engine is strengthened by the circumstantial evidence that the outflow-borne
masers all lie outside the shadow cast by the truncated disk.  This
would not be true for a broader disk.  However, if the outflow subtends almost
90 \% of $4\pi$ str as seen from the central engine, then the accretion disk
may be starved such that it will be exhausted on the order of $10^7$ yr,
assuming a mass of $10^5$ M$_\odot$, 10\% accretion efficiency, and 10\%
emission fraction for hard X-rays.

The position angles at which the outflow is free of occultation by
the 0.4 pc radius disk coincide with the limits of the kiloparsec
scale ionization cone observed west of the
nucleus (\cite{vb97}).  The southern edges of the redshifted maser outflow and
the ionization cone, both lie at position angles of $\sim -120^\circ$. The 
northern edge of the maser outflow, at  $16^\circ$, coincides roughly with the 
position angle of the northernmost [O\,III] filament ($\sim -20^\circ$) or a
northeastern blue knot ($\sim 10^\circ$). In addition, the mean axis of the
maser outflow ($-52^\circ$), corresponds well to the orientation of the
dominant [O\,III] filament  ($\sim -50^\circ$) and a radio hotspot, observed
with arcsecond resolution (\cite{elmouttie98}), which could mark a radio jet,
though none has yet been observed directly. 
 

This work has been conducted in collaboration with R. S.
Booth$^{(1)}$, S. P. Ellingsen$^{(2)}$, J. R. Herrnstein$^{(3)}$,  D. L.
Jauncey$^{(4)}$, P. M. McCulloch$^{(2)}$, J. M. Moran$^{(5)}$, R. P. Norris,
J. E. Reynolds, and A. K. Tzioumis$^{(4)}$.

\hrulefill

$^{(1)}$Onsala Space Observatory; $^{(2)}$ University of
Tasmania; $^{(3)}$Renaissance Technologies; $^{(4)}$Australia Telescope
National Facility; $^{(5)}$Harvard-Smithsonian Center for Astrophysics

\begin{figure}[!h]
\centerline{\psfig{figure=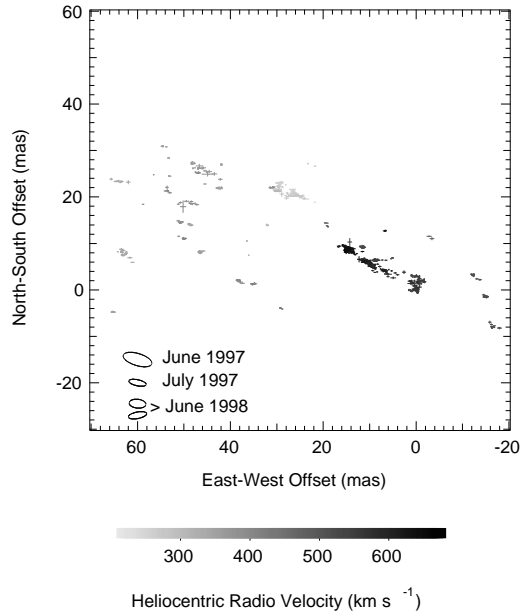,width=7cm}
}
\caption{Sky distribution of H$_2$O maser emission at three epochs
(superposed). The synthesized beam for each epoch is indicated. Symbol shading
indicates line-of-sight velocity for each spectral channel plotted.   Error bars
indicate total position uncertainties ($1\sigma$).}
\end{figure}

\begin{figure}[!h]
\centerline{\psfig{figure=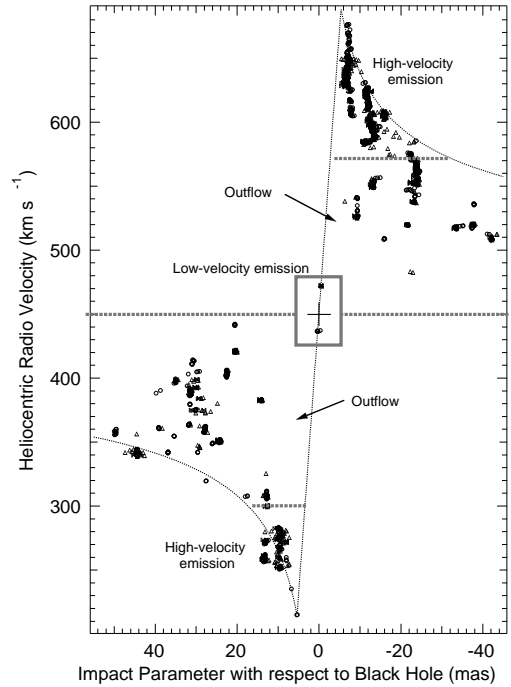,width=7cm}
}
\caption{Position-velocity diagram annotated to highlight the proposed disk and
outflow components.  Dashed rotation curves correspond to the midline of the
disk.  The steep diagonal line corresponds to the near side of the disk, at the
inner radius.  The outflow component is Doppler shifted with respect to the
systemic velocity by on the order of
$\pm100\kms$. }
\end{figure}

\begin{figure}[!h]
\centerline{\psfig{figure=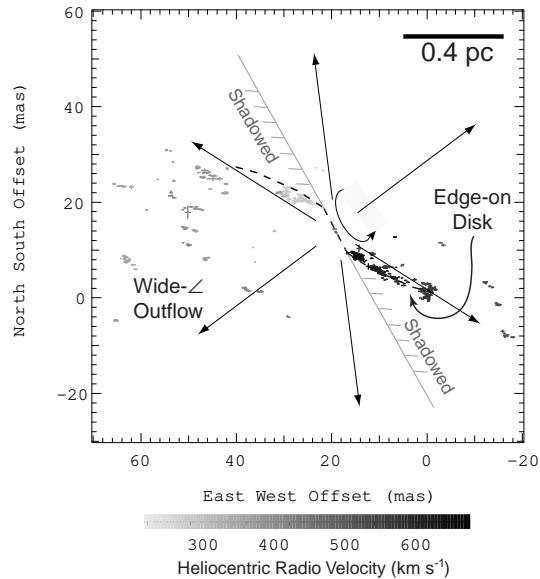,width=7cm}
}
\caption{Model of the  disk-outflow structure in the inner 1 pc of the AGN
overlayed on the maser emission distribution. Dashed lines indicate a warped edge-on
disk.  Outward facing arrows indicate a wide-angle outflow.   Regions
without a direct line of sight to the central engine are shadowed by the
disk. 
 }
\end{figure}

\end{document}